\documentclass[letterpaper]{article}
\usepackage{iccc}

\usepackage{anonymize} 		   

\usepackage{times}
\usepackage{helvet}
\usepackage{courier}
\usepackage{footmisc}
\usepackage{url}
\makeatletter
\g@addto@macro{\UrlBreaks}{\UrlOrds}
\makeatother
\usepackage{amsmath}
\usepackage[normalem]{ulem}
\useunder{\uline}{\ul}{}
 \newcommand{\etal}[0]{\textit{et al.}}
 
\title{Can GAN originate new electronic dance music genres?---Generating novel rhythm patterns using GAN with Genre Ambiguity Loss}

\author{
		Nao Tokui\\
       \small{Graduate School of Media and Governance}\\
       \small{Keio University}\\
       \small{5332 Endo, Fujisawa}\\
      \small{Kanagawa, Japan}\\
       \small{tokui@sfc.keio.ac.jp}
} 
       
\begin{document}

\maketitle

\begin{abstract}
\begin{quote}
Since the introduction of deep learning, researchers have proposed content generation systems using deep learning and proved that they are competent to generate convincing content and artistic output, including music. However, one can argue that these deep learning-based systems imitate and reproduce the patterns inherent within what humans have created, instead of generating something new and creative. 

In this paper, we focus on music generation, especially rhythm patterns of electronic dance music, and discuss if we can use deep learning to generate novel rhythms, interesting patterns not found in the training dataset.  

We extend the framework of Generative Adversarial Networks(GAN) and encourage it to diverge from the inherent distributions in the dataset by adding additional classifiers to the framework. The paper shows that our proposed GAN can generate rhythm patterns that sound like music rhythms but not belong to any genre in the training dataset.

The source code, generated rhythm patterns, and a supplementary plugin software for a popular Digital Audio Workstation software are available on our website\footnote{\url{https://cclab.sfc.keio.ac.jp/projects/rhythmcan/} \url{https://github.com/sfc-computational-creativity-lab/x-rhythm-can}\label{main_website}}.

\end{quote}
\end{abstract}

\section{Introduction}

Since the research of deep learning took off, researchers and artists have proved that deep learning models are competent for generating images,  texts, and music. Many researchers have been working on applications of deep learning, especially architectures for time-series prediction such as Recurrent Neural Networks(RNNs), in music generation\cite{Briot}.

For image generation, the framework of {\textit Generated Adversarial Networks(GAN)} is the most popular one among deep learning-based generative models\cite{Goodfellow2014a}. Since its introduction in 2014, the qualitative and quantitative quality of images generated by GAN has steadily improved. Researchers have shown that GAN is capable of generative photo-realistic images of human faces\cite{Karras2019}. 

In this paper, we propose to use an extended GAN model to generate new interesting rhythm patterns. The original GAN framework consists of two neural networks; the generator and the discriminator. The objectives of these networks are as follows: 

\begin{description}
\item The generator, $G$: To transform random noise vectors ($z$) into "fake" samples, which are similar to real samples (i.e., images) drawn from the training dataset.
\vspace{1mm}
\item The discriminator, $D$: To estimate the probability that a sample came from the training data rather than from $G$.
\end{description}

GAN can be formalized mathematically in a min-max formula: 

\begin{align}
\min_{G}\max_{D}V(G, D) = \log(D(x)) + \log(1-D(G(z)))
\label{gan_func}
\end{align}

\begin{description}
\item $V(G,D)$: the objective, which $D$ maximizes, and $G$ minimizes.

\item $D(x)$: the probability that input x came from the real data according to D

\item $G(z)$: the output of $G$ with $z$ random input noise
\end{description}

\begin{figure}[bt]
  \includegraphics[width=1\columnwidth]{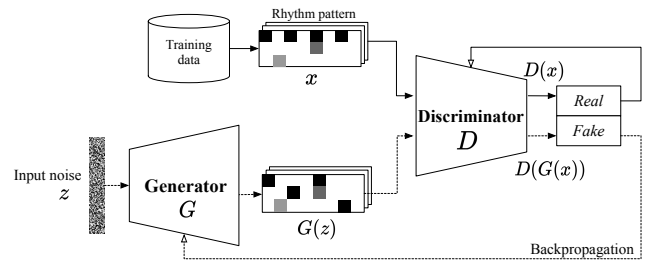}
  \caption{Overview of Generative Adversarial Networks(GAN)}
  \label{fig:gan}
\end{figure}

The objective of $D$ is to maximize $D(x)$ and $(1-D(G(z))$ terms, so it is trained to distinguish real data from synthetic data correctly. The $G$'s objective, on the contrary, is to minimize $1-D(G(z))$ term, i.e., to maximize D(G(z)) term. The $G$ is trained to fool $D$ and make $D$ to believe the synthetic images are real ones drawn from the dataset. 

The GAN framework assumes that this min-max setting leads to better $G$ and $D$; eventually, $G$ will be able to generate data that has a similar distribution as one in the training data and fool $D$. 

Although GAN's ability to generate high-quality images is proved and it is one of the unsupervised learning methods, instead of supervised learning, still one can argue that GAN merely learns the underlying patterns of the training data and reproduces them. In other words, GAN is not formulated to generate anything fundamentally novel and original.  

Let's assume that we trained a GAN model on images of classical painting. Since G is trained to generate images that fool D, in theory, eventually, G will just generate images that look like already existing paintings. Although G may accidentally create something new by mixing/generalizing the existing painting, G is not incentivized to explore the creative edge. 

Theoretically speaking, the most desirable G is the one that generates every single image in the dataset and nothing else, so that D will be completely fooled by the G. Hence we cannot expect Gs to be creative in the formulation of GAN. 

The question of originality is not only an artistic one but also an economic and social one. If the model simply copies the training data, music generated by the model can infringe on the copyright of the original music in the dataset (unless they are in the public domain). 


To tackle this problem of originality, the author proposes to extend the GAN framework for music generation by adding the second $D$ and introduce \textit{Genre Ambiguity Loss}. The second $D$ is a classification model to classify the genre of generated music. Genre Ambiguity Loss quantifies how ambiguous the genre classification is and penalizes if the classification is too confident; the loss is minimum when the output of the classification model is equiprobable. In this way, the $G$ is incentivized to generate music in a new style(genre), while the music has a somewhat similar statistical distribution as the entire training data.

Elgammal \etal introduced this GAN setting in \cite{Elgammal2017} and showed the extended GAN model trained with a dataset of historical paintings in various styles(such as impressionism, cubism) could generate abstract paintings that look like paintings, but at the same time, do not belong to any of genres in the dataset. 

The domain of music this paper handles is the rhythm patterns of electronic dance music. From drum and bass(DnB), dubstep to trap and juke, electronic dance music is one of a few music genres, in which new sub-genres emerge every couple of years, and often, the rhythm is the key factor that characterizes these sub-genres. Hence the question this paper poses can be boldly summarized as follows: can we use AI, more specifically GAN,  to originate new electronic dance music genres? 

\section{Related Works}

\subsection{Music Generation with GAN}

As compared to image generation, researches applying GAN to music generation are relatively few. 

Briot \etal conducted a comprehensive survey on deep learning-based music generation\cite{Briot}. Out of 32 papers the survey covered, only two researches used some sort of GAN architecture in their papers(Chapter 7).

C-RNN-GAN is one of the earliest examples of GAN-based music generation\cite{Mogren2016}. The objective of C-RNN-GAN is to generate a single voice polyphonic music of classical music.  Both the generator and the discriminator are 2-layer LSTM architecture\cite{Hochreiter1997}. The difference between them is that the LSTM in the discriminator is bidirectional so that it can take both past and future patterns into account.  

Li-Chia Yang \etal proposed MidiNet architecture to generate pop music melodies\cite{Yang2017}. The architecture quantizes melodies in the training dataset at the 16th notes and represents them in piano roll like format. The biggest difference between C-RNN-GAN and MidiNet is that MidiNet utilizes convolutional layers both for the discriminator and the generator.  The paper also proposed an extended version of MidiNet, which takes a one-hot vector for the chords, specifying its key and if it's minor or major in addition to the random $z$ vector of the GAN architecture.

\subsection{Rhythm Generation}

Historically speaking, the majority of researches on music generation deal with melodies and harmonies. In the same survey\cite{Briot}, only two papers handle rhythm as their main objective. However, researchers started experimenting with rhythm generation techniques recently. 

\citeauthor{Choi}(\citeyear{Choi}) showed that a simple LSTM architecture trained with rhythm data encoded as strings of characters could successfully generate heavy-metal drum patterns. The paper used a binary representation of nine different drum components. For example, 100000000 and 010000000 represent kick and snare, respectively, and 110000000 means playing kick and snare simultaneously.  

Markris \textit{et al.}\shortcite{Makris} proposed an architecture with stacked LSTM layers conditioned with a Feedforward layer to generate rhythm patterns.
The Feedforward layer takes information on accompanying bass lines and the position within a bar. Training rhythm patterns in binary representation, similar to \cite{Choi}, are fed into LSTM layers. The system allows only five drum components (kick, snare, tom, hihat, cymbal).   

In \cite{Gillick2019}, Gillick \textit{et al.} employed Variational Autoencoder(VAE) model to generate new rhythm patterns. The paper also made use of the encoder-decoder architecture of VAE and proved that their proposed architecture could "humanize" quantized rhythms by slightly shifting the timing, and stressing or weakening certain onsets. 

Their model handles a rhythm pattern represented as three matrices: onsets, velocities (strength of onsets), timing offsets from quantized timing. The VAE model consists of layers of bidirectional LSTM used in \cite{Roberts2018b}. In this paper, we adopted a similar data representation but only dealt with the matrix of the velocities.

\subsection{GAN and Creativity}

It has been proved that GANs can generate photo-realistic images of human faces\cite{Karras2019}.  Many artists started experimenting with GAN as a means to generate visual materials, and an artist group even managed to sell their GAN-generated "painting" at one of the most prestigious auction houses in the world\footnote{Is artificial intelligence set to become art’s next medium? \url{https://www.christies.com/features/A-collaboration-between-two-artists-one-human-one-a-machine-9332-1.aspx}}. As the research on the GAN technique progresses, the quality of the generated images—both in image resolution and realisticness— are getting better and better; however, the originality of them is open to question.

Is the GAN algorithm creating something novel? Or does it just memorize the inherent patterns within the dataset and learn to reproduce them? 

Elgammal \etal\shortcite{Elgammal2017} tackled this problem by extending the GAN framework to have the second \textit{style} discriminator and named it \textit{Creative Adversarial Networks(CAN)}. They showed that CAN could generate unique abstract paintings based on the dataset of historical paintings in various styles (Impressionism, Cubism, etc.)

The discriminator of normal GAN (GAN discriminator) is trained to differentiate generated data(painting, in this case) from the original data in the training dataset. CAN framework has the second disseminator, which is a classification model to classify the generated image into one of the established styles in training data.

The generator is trained to not only fool the GAN discriminator by generating images that look like paintings but also to puzzle the style discriminator, i.e., the generator gets trained to generate images that make the classification result of the style discriminator as equiprobable as possible. The authors named this loss {\it Style Ambiguity Loss}.  

The first discriminator makes sure that the generated image looks like a painting, but at the same time, the second discriminator(style discriminator) pushes the painting diverting from the established styles. These two contradicting forces make it possible for CAN to learn to generate unique realistic paintings that don't belong to the styles in the past. 

A more precise definition will follow in the next section.

\section{Proposed Method}
%
%
%
%
%

\subsection{Genre-conditioned GAN}

\begin{figure}[bt]
\centering
  \includegraphics[width=1\columnwidth]{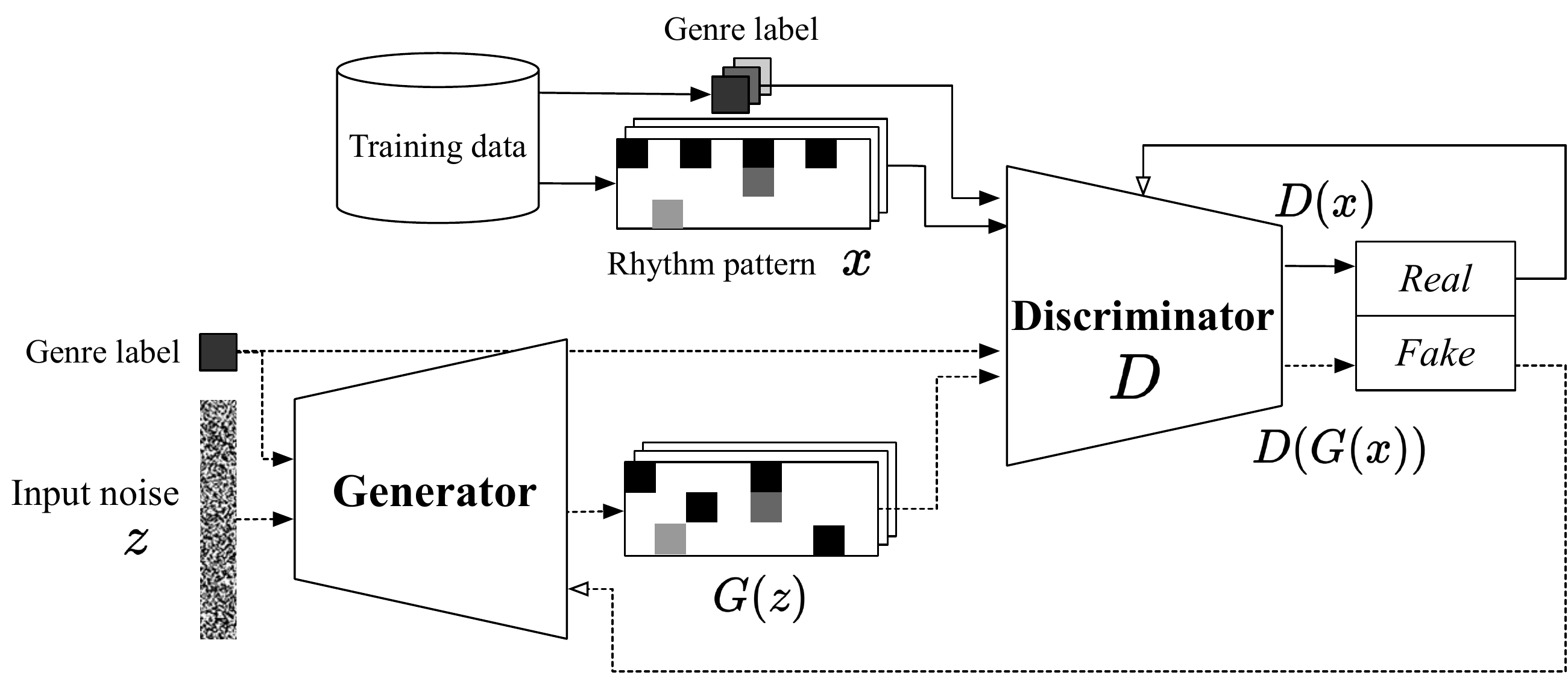}
  \caption{Overview of Genre-conditioned GAN}
  \label{fig:conditioned-gan}
\end{figure}

At first, the author tested a GAN architecture conditioned on the music genre to see whether or not the proposed GAN model can generate rhythm patterns in a specified genre(Fig.\ref{fig:conditioned-gan}).

In this genre-conditioned setting, the discriminator and the generator take an additional input $y$ to condition on the genre. $y$ is a scalar number, $[0, K)$, where $K$ is the number of genres in training data. 

In $D$, the input $y$ is converted into a dense vector in the same shape as the drum onset matrix described in the Data Representation section, through an embedding layer. The output of the embedding layer is concatenated with the drum onset matrix and fed into two layers of bidirectional LSTM.  

$G$ embeds $y$ into a vector of the same size as $z$  and then multiplies the embedded label and $z$ element-wise. The resulting joined representation is used as input of the generator architecture.  Then $G$ and $D$ train adversarially with drum onset data and the corresponding genre labels. These architectures are based on \cite{Mirza2014}.

\subsection{GAN with Genre Ambiguity Loss - Creative-GAN}

\begin{figure*}[bt]
\centering
  \includegraphics[width=0.8\textwidth]{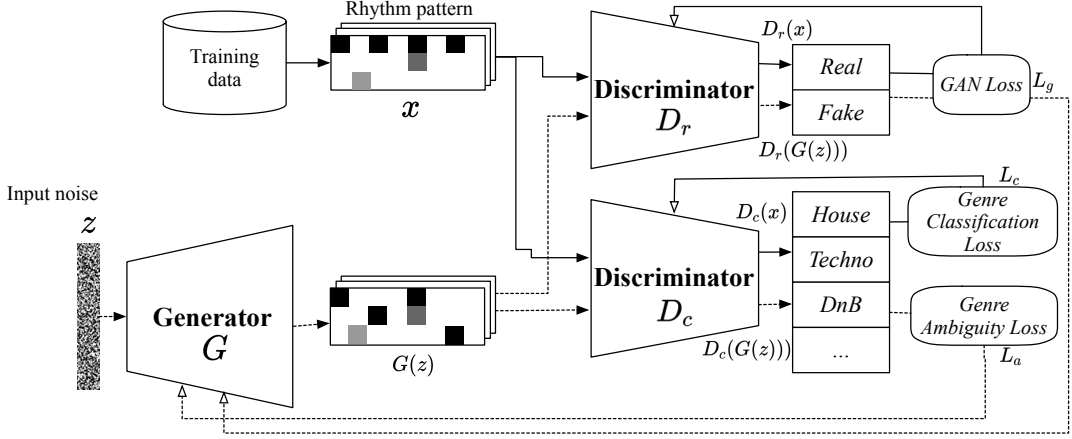}
  \caption{Overview of GAN with Genre Ambiguity Loss}
  \label{fig:can}
\end{figure*}

In the second experiment, the GAN architecture doesn't have the genre conditioning input; instead, it has the second $D$ in addition to the original $D$. Here, we note the original GAN $D$ differentiating the real drum patterns from the generated ones as $D_r$, and the second $D$ classifying genres as $D_c$. We refer to this GAN setting as \textbf{\textit{Creative-GAN}} in short. 

While training the discriminator, $D_c$ is trained to classify $K$ genres of real and generated drum patterns.  In other words, we added \textit{Genre Classification Loss}, $L_c$, to the cost function of the discriminator $D$.

On the other hand, $G$ is encouraged to \textit{confuse} both $D_r$ and $D_c$; we train $G$ to not only make $D_r$ to believe generated drum patterns are real---drawn from training data---, but also make $D_c$ to be uncertain about the genres of the generated patterns. 

The entropy of class posteriors of generated patterns is maximized, when the class posterior, $p(c|G(z))$, is equiprobable. Similarly, the cross-entropy between the class posterior and uniform distribution is minimized,  when the class posterior is equiprobable. In this experiment, we chose to encourage $G$ to minimize this cross-entropy, instead of maximizing the entropy of class posteriors. 

We use this cross-entropy as the additional loss corresponding to the uncertainty of $D_c$, and call it \textbf{\textit{Genre Ambiguity Loss}}(\textit{Style Ambiguity Loss} in \cite{Elgammal2017}). 

The entire cost function with adversarial objectives of $G$ and $D$ can be redefined as follows:

\begin{align}
\begin{split}
\min _{D} \max _{G}=  \log \left(D_{r}(x)\right)+\log \left(1-D_{r}(G(z))\right) \\
+\log \left(D_{c}(c=\hat{c} | x)\right)\\
-\sum_{k=1}^{K}\left(\frac{1}{K} \log \left(D_{c}\left(c_{k} | G(z)\right)\right) \right. \\ 
\left. +\left(1-\frac{1}{K}\right) \log \left(1-D_{c}\left(c_{k} | G(z)\right)\right) \vphantom{\frac{1}{K}} \right)
\label{func:can}
\end{split}
\end{align}

\noindent where $x$ and $\hat{c}$ are a real rhythm pattern its genre label drawn from the training data distribution $p_{data}$. $z$ is a random noise input vector, so $G(z)$ denotes generated rhythm patterns, and $K$ denotes the number of genre classes in training data.

\begin{description}

\item Discriminator:  In function (\ref{func:can}), $D$ is encouraged to maximize $\log (D_r(x))+\log (1-D_r(G(z)))$ by differentiating properly real rhythm patterns from fake (generated) ones---$D_r$ should output 1 for real ones and 0 for fake.  $D$ is also encouraged to maximize $\log D_{c}(c=\hat{c} | x)$ by classifying the genre of real rhythm patterns; one can think $D_c$ as a simple classification model with $N_c$ class labels.

\item Generator: $G$ is encouraged to minimize $\log (1-Dr(G(z)))$ by fooling $D_r$. $G$ is also encouraged to maximize $-\sum_{k=1}^{K}\left(\frac{1}{K} \log \left(D_{c}\left(c_{k} | G(z)\right)\right)\right)+\left(1-\frac{1}{K}\right) \log \left(1-D_{c}\left(c_{k} | G(z)\right)\right)$, multi label cross-entropy loss with $K$ number of classes. As mentioned earlier, this loss can be maximized when the cross-entropy with uniform distribution is minimized. 

\end{description}

\section{Experiments}
\subsection{Dataset}

As a training dataset of electronic music rhythm patterns,  the author used a MIDI pattern library commercially available.  Groove Monkee Mega Pack GM\footnote{\url{https://groovemonkee.com/products/mega-pack}} includes 34828 MIDI files of rhythm patterns of various music genres. These genres include rock, blues, country, and electronic music.  

Out of 34828, 1454 MIDI files are labeled as electronic and in the generic General MIDI(GM) format, and electronic music consists of sub-genres: \textit{Breakbeats, DnB(drum and bass), Downtempo, Dubstep, Four To The Floor, Garage, House, Jungle, Old Skool(Hip Hop), Techno, Trance}.  In this experiment,  the author excluded genres that came with the fewer number of MIDI files, namely \textit{Dubstep and Four to The Floor},  and used the nine other genres(Table.\ref{table:genres}). $K$ denotes the number of genres($K = 9$). 

\begin{table}[]
\centering
\begin{tabular}{c}
Breakbeats         \\
DnB(drum and bass) \\
Downtempo          \\                           
Garage             \\                           
House			   \\
Jungle			   \\
Old Skool		   \\
Techno 			   \\
Trance 			   \\
\end{tabular}
\caption{Electronic dance music genres in training data}
\label{table:genres}
\end{table}

\begin{table}[]
\centering
\begin{tabular}{lll}
MIDI Note \# & Instrument     & \# of Onsets \\
42           & Closed Hi Hat  & 34260        \\
38           & Acoustic Snare & 20691        \\
36           & Bass Drum 1    & 19803        \\
46           & Open Hi-Hat    & 5069         \\
44           & Pedal Hi-Hat   & 3807         \\
51           & Ride Cymbal 1  & 1494         \\
37           & Side Stick     & 936          \\
56           & Cowbell        & 602          \\
45           & Low Tom        & 514          \\
47           & Low-Mid Tom    & 500          \\
48           & Hi-Mid Tom     & 440          \\
43           & High Floor Tom & 308          \\
41           & Low Floor Tom  & 72           \\
39           & Hand Clap      & 62           \\
40           & Electric Snare & 42           \\
53           & Ride Bell      & 35           \\
49           & Crash Cymbal 1 & 13           \\
57           & Crash Cymbal 2 & 1           
\end{tabular}
\caption{Total numbers of onsets of each drum instrument}\label{table:total_onsets}
\end{table}

\begin{table}[]
\centering
\begin{tabular}{cr}
\textbf{Instrument} & \textbf{\# of Onsets} \\ \hline
Hi-hat closed       & 34260                 \\
Snare               & 20733                 \\
Kick                & 19803                 \\
Hi-hat open 88      & 8876                  \\
Cymbal              & 1508                  \\
Low Tom             & 1086                  \\
Rim                 & 936                   \\
High Tom            & 748                   \\
Clap/Cowbell        & 699                  
\end{tabular}
\caption{Total numbers of onsets of each drum instrument (after mapping into the selected nine drums)}\label{table:mapped_onsets}
\end{table}

Table.\ref{table:total_onsets} shows the number of onsets---drum hits--- of each drum instrument in the MIDI files. These drum instruments are mapped into nine different drums, namely \textit{Kick, Snare, Hi-hat, Cymbal, Low Tom, High Tom, Clap/Cowbell, Rim}, in the experiments throughout this paper. The author picked the number nine following previous literatures\cite{Gillick2019}. Table.\ref{table:mapped_onsets} shows the distribution of the total number of onsets in the MIDI files.

\begin{figure}[hbt]
  \includegraphics[width=1\columnwidth]{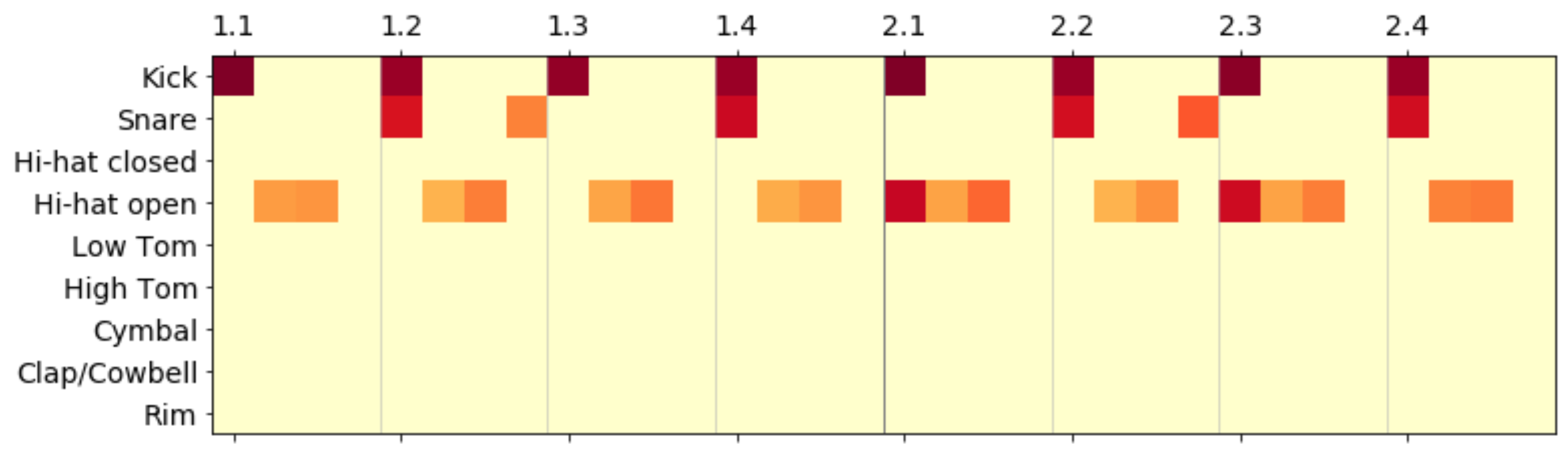}
  \includegraphics[width=1\columnwidth]{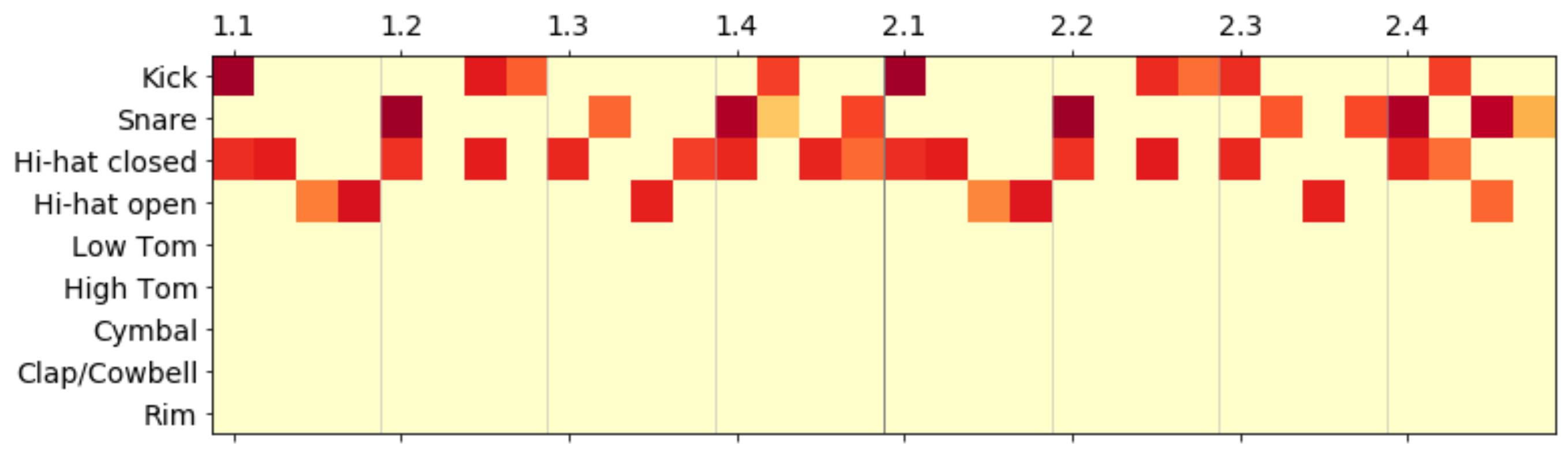}
  \caption{Examples of the rhythm pattern in training data represented as onset matrices(top: \textit{House}, bottom: \textit{Breakbeats})}
  \label{fig:onset_matrix}
\end{figure}

\subsection{Rhythm Similarity/Dissimilarity}\label{sec:rhythm-similarity}

\begin{figure}[bt]
  \includegraphics[width=1\columnwidth]{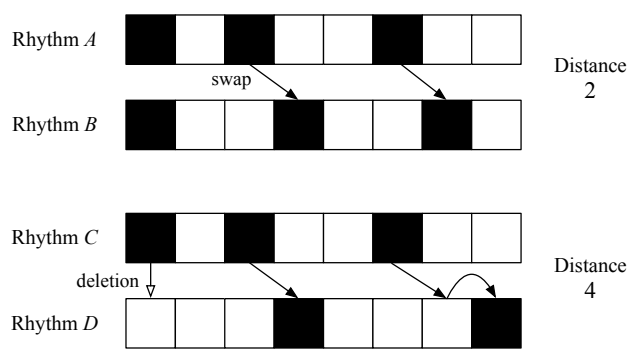}
  \caption{Swap/Edit distance between two rhythm patterns}
  \label{fig:rhythm_distance}
\end{figure}

Toussaint\shortcite{TOUSSAINT2006} proposed and examined various metrics for quantifying the similarity and dissimilarity between different rhythm patterns. The paper showed that "swap distance" is the most effective among the metrics he tested. The swap distance between two rhythm patterns is the minimum number of swaps required to convert one rhythm to the other.  In Fig.\ref{fig:rhythm_distance}, the swap distance between Rhythm A and B is two because it requires two swaps to convert A to B. 

Swap distance assumes strings in the same length; in this case, rhythm patterns with the same number of onsets. Therefore we used the generalized "edit distance" to calculate the distance between rhythm patterns with different numbers of onsets. Edit distance allows three operations: insertion, deletion, and swap, so that it can handle patterns with different lengths(Fig.\ref{fig:rhythm_distance} bottom). While calculating the distance, we ignored the velocity of onsets.

As mentioned earlier, we are dealing with nine different drum instruments within a rhythm pattern. Hence, we summed up edit distances of 9 pairs of patterns for each instrument and used it as a distance of the given two drum patterns.  

\begin{figure}[bt]
  \includegraphics[width=1\columnwidth]{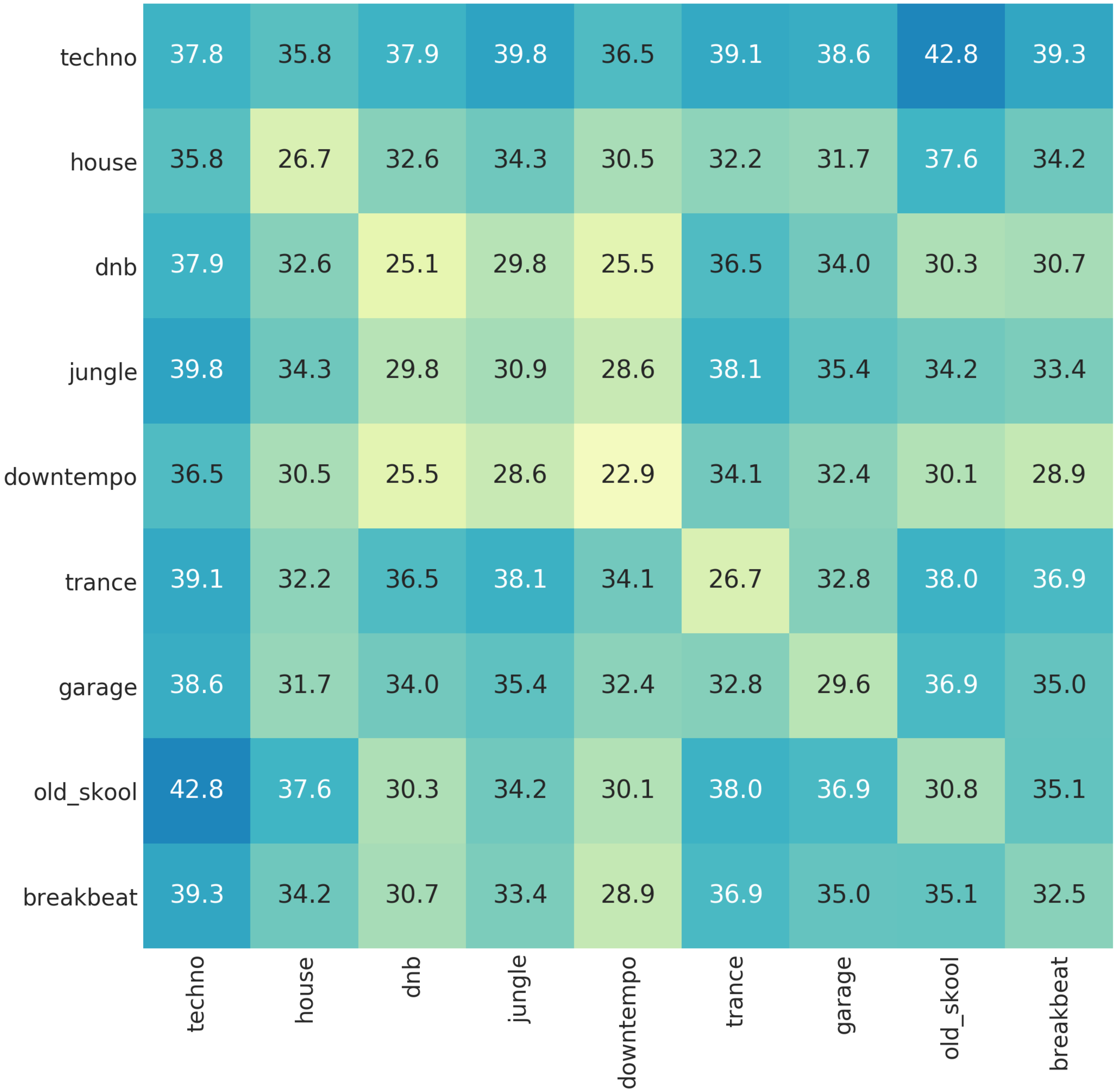}
  \caption{Distance Matrix among rhythm patterns of various genres in the dataset}
  \label{fig:dataset_distmatrix}
\end{figure}

As a preliminary test, the author calculated edit distances of all drum patterns in the dataset and categorized them with genres of the patterns.  Fig.\ref{fig:dataset_distmatrix} shows the average distances between rhythm patterns in each genre. 

It shows that Downtempo, DnB, House and Trance are the most homogenous genres in terms of rhythm patterns. It also shows Downtempo, DnB, and Jungle have something in common. The rhythm in these genres is based on breakbeats, and the main difference between Downtempo and the other two is mainly how fast and densely these rhythms are played. The fact that DnB derived from Jungle explains the similarity between the two. On the other hand, Old Skool (mainly old school HipHop beats) and Techno appear to be the most different.

\subsection{Genre Classification}

As the second preliminary test, the author trained a classification model that classifies all genres based on rhythm patterns. The classification models are similar to the discriminator in our GAN experiments described later. The model consist of two layers of Bidirectional LSTM. 
The following section describes the representation of rhythm patterns in detail.

The experiments showed that the model could distinguish the genres with more than 90\% validation accuracy($93.09\%$).  

The high accuracy means that it is sometimes difficult for us to differentiate rhythm patterns in similar genres (i.e., Breakbeat and Downtempo), there are differences among them, which the model can exploit and classify correctly.

\subsection{Data Representation}\label{sec:datarep}

The models proposed in this paper handle two-bar-long rhythm patterns. A rhythm pattern is represented as a matrix of drum onset with velocity ---strength--- with a fixed timing grid (Fig.\ref{fig:onset_matrix}). 

The unit of the timing grid ---the minimum time step of drum onset--- is the 16th note; the timing of every drum onset is quantized to the timing of the nearest 16th note. Each onset is represented as $[0., 1.]$ by normalizing the corresponding MIDI velocity values $[0, 127]$ to $[0., 1.]$. 

Since onset timings of two-bar-long patterns are quantized in 16th notes, and all drums are mapped into the nine instruments, all matrices are $9 \times 32$. We'll call this matrix "onset matrix" in this paper.  

\subsection{Implementation}

$D$ of the both Genre Conditioned-GAN and the Creative-GAN with Genre Ambiguity Loss uses two layers of Bidirectional LSTM with 64 nodes. The output of LSTM layers is fed to two layers of fully-connected dense layers. The last layer has one node with Sigmoid activation function to tell if the input pattern is real(training data) and fake(generated). In Creative-GAN, $D$ has the second output with $K$ nodes with Softmax activation and categorical cross-entropy loss. 


$G$ in both GAN models has three layers of LSTM with (128, 128, 9) nodes. The last layer outputs a batch of $9 \times 32$ matrix, which is the same size as the onset matrix described in the last section. The dimensionality of random input $z$ for $G$ is 100. This $z$ input is fed into two layers of dense layers and reshaped to the format compatible with LSTM layers. We used LeakyReLU as an activation function, except for the last layer, and Adam Optimizer. 

We implemented these models using Keras framework with TensorFlow backend. The source code and training data in NumPy format are available online\footref{main_website}. 

\section{Results}

\subsection{Conditional GAN}

The first experiment showed that the proposed GAN architecture conditioned on a genre could generate rhythm patterns in the specified genre. You can listen to examples of generated rhythm patterns on our website\footref{main_website}.  

\begin{figure}[bt]
  \includegraphics[width=1\columnwidth]{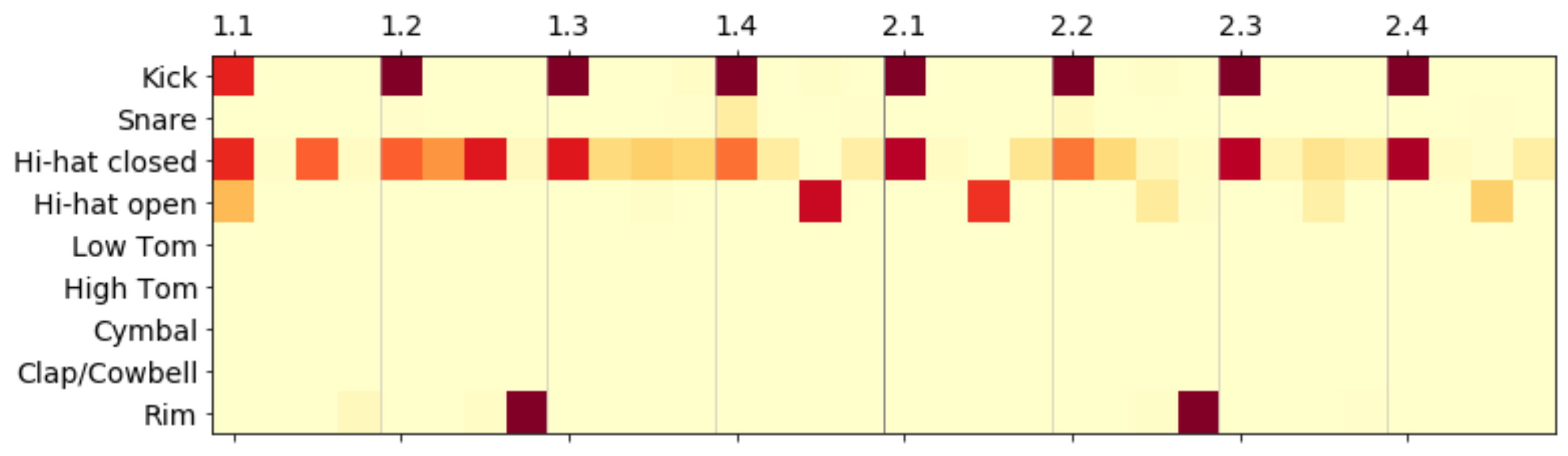}
  \includegraphics[width=1\columnwidth]{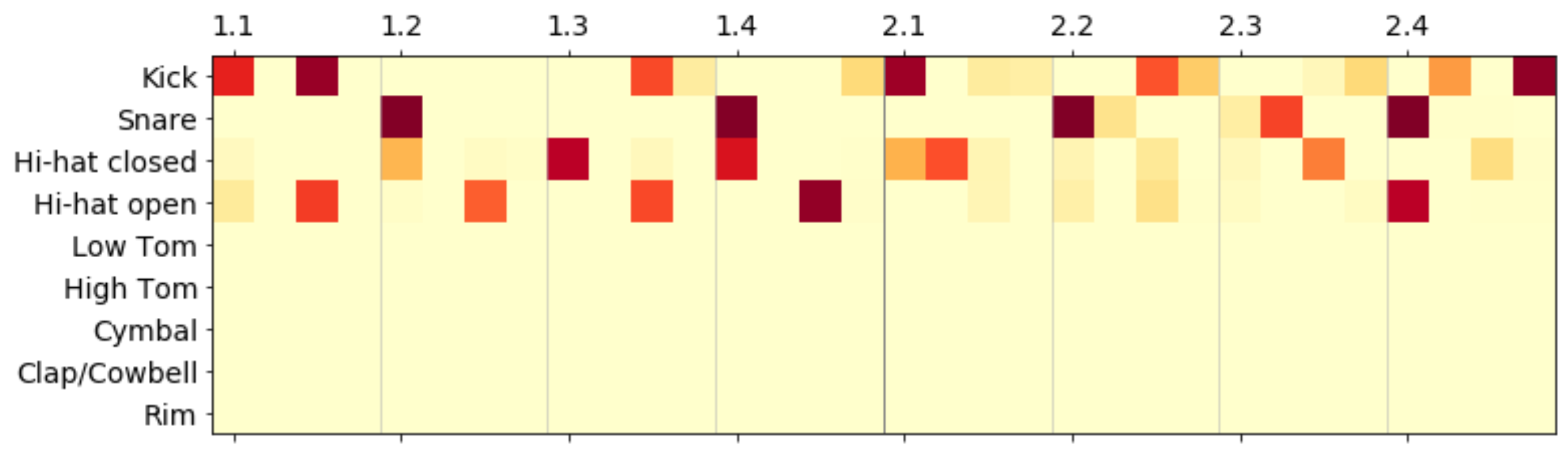}
  \caption{Examples of rhythm patterns generated by the trained Genre-conditioned GAN model (Top: \textit{House}, Bottom: \textit{Breakbeats})}
  \label{fig:generated_conditioned}
\end{figure}

Fig.\ref{fig:generated_conditioned} shows some of the examples of generated rhythm patterns and corresponding genre labels. 



\subsection{GAN with Genre Ambiguity Loss - Creative-GAN}

\begin{figure}[bt]
  \includegraphics[width=1\columnwidth]{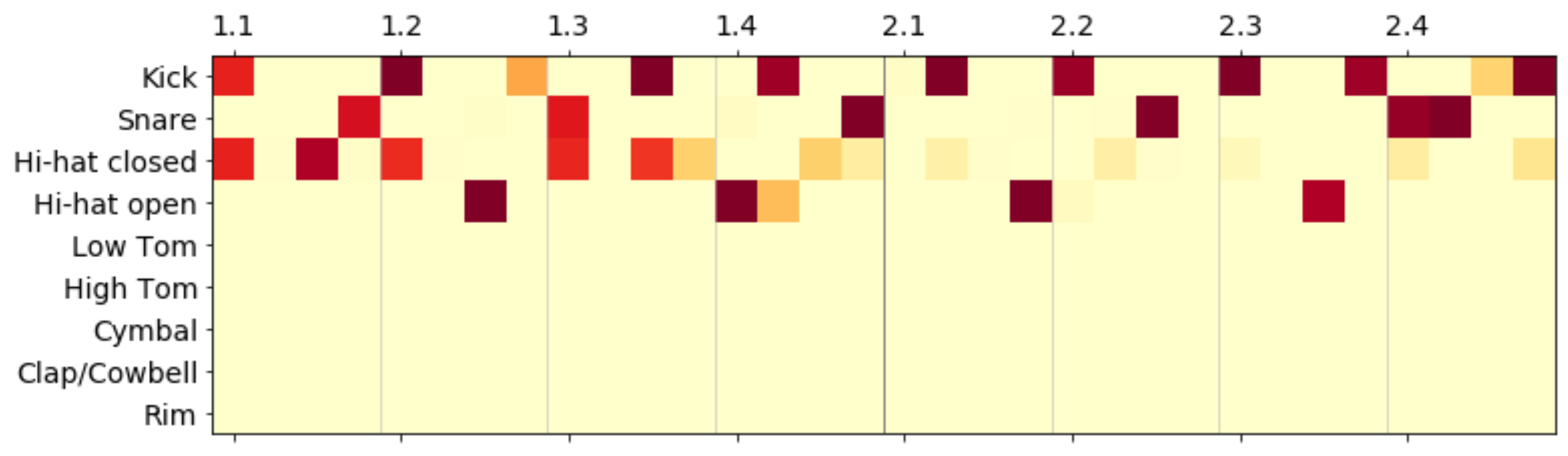}
  \includegraphics[width=1\columnwidth]{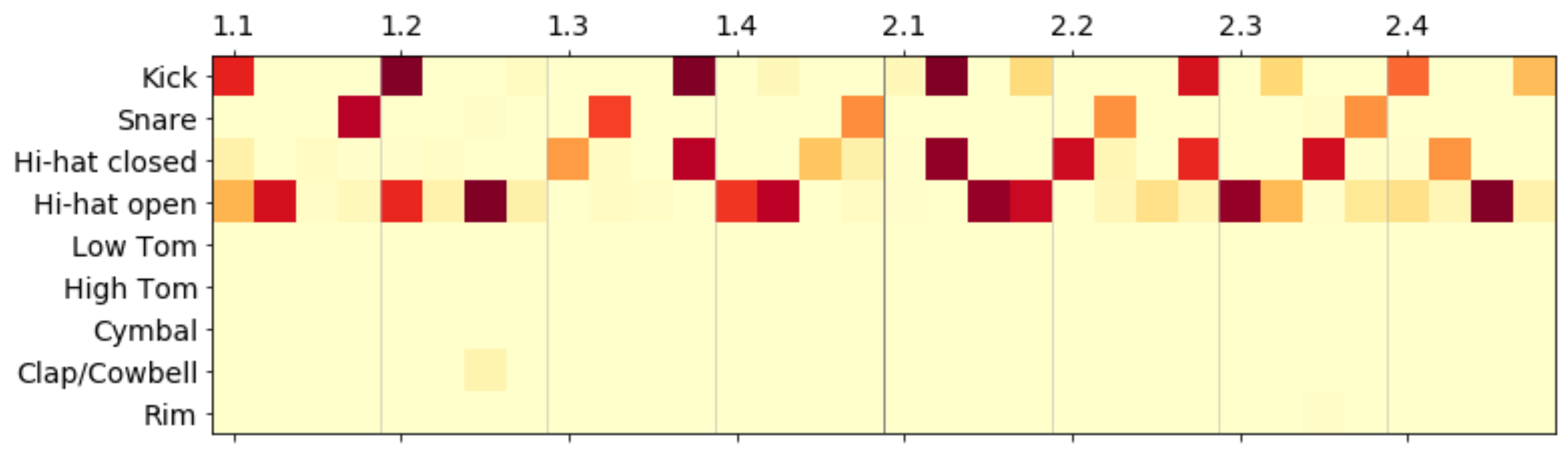}
  \includegraphics[width=1\columnwidth]{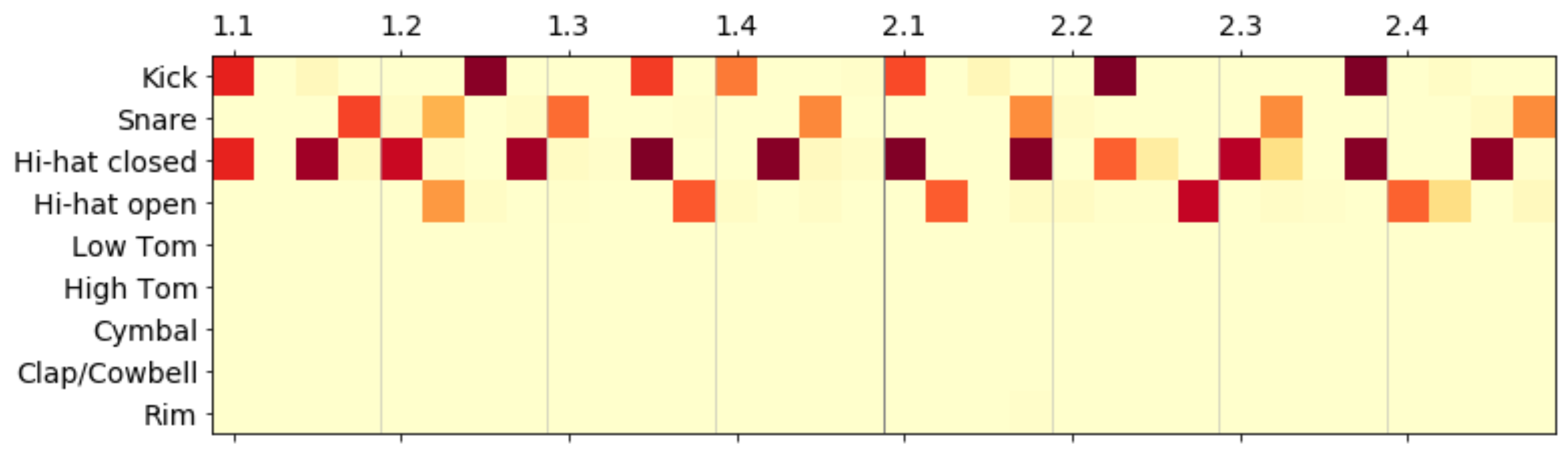}
  \includegraphics[width=1\columnwidth]{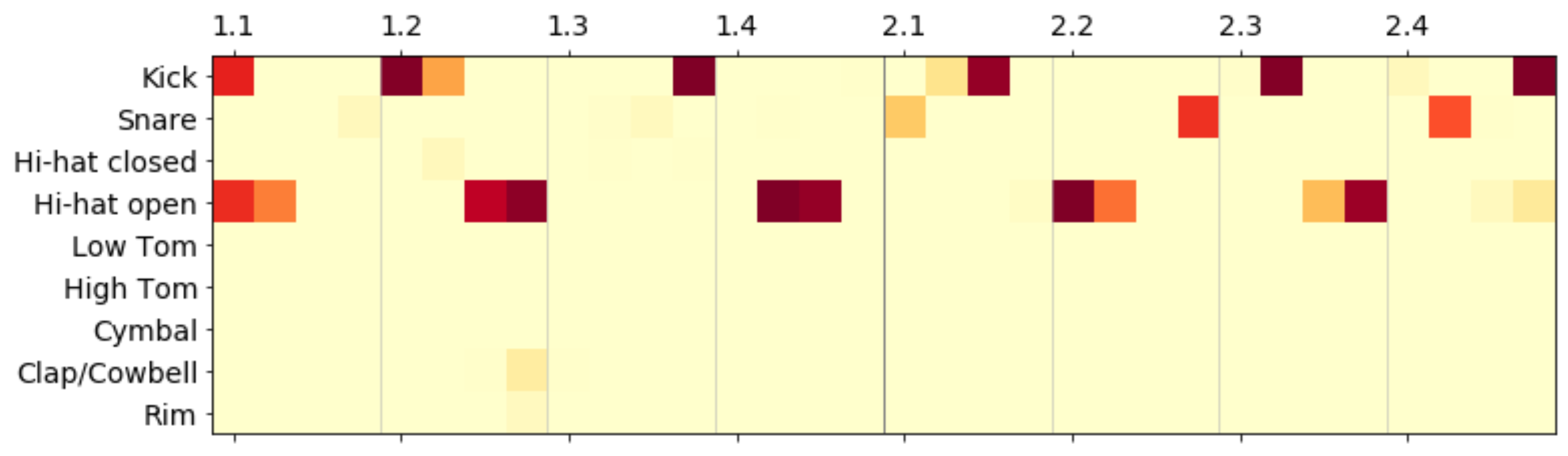}
  \caption{Examples of rhythm patterns generated by the trained Creative-GAN model}
  \label{fig:generated_cgan}
\end{figure}

The second model with Genre Ambiguity Loss---Creative-GAN model--- also worked as expected. The training of the model converged around 90 epochs and did not get any better. Fig.\ref{fig:generated_cgan} shows some of the examples of generated patterns by the generator from the 94th epoch. Again you can listen to these examples on our website.

\begin{figure}[bt]
  \includegraphics[width=1\columnwidth]{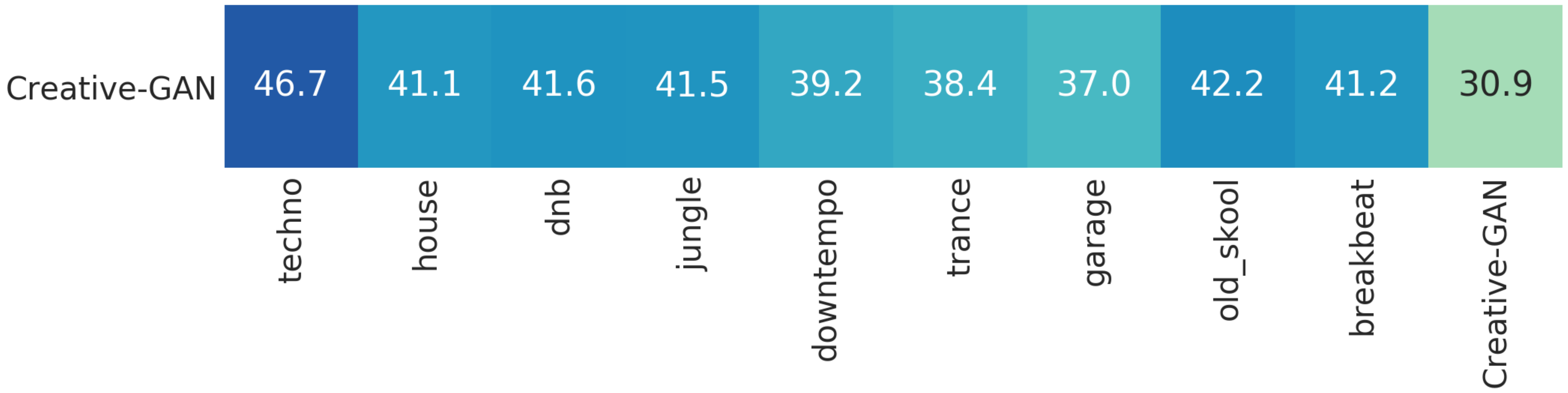}
  \caption{Average distances between rhythm patterns generated by the trained Creative-GAN model and patterns in training data}
  \label{fig:distance_matrix_cgan}
\end{figure}

We generated 500 patterns with random $z$ input vectors and calculated the distances from the training data. Fig.\ref{fig:distance_matrix_cgan} shows high distances for all of the genres in the training data. It means that the generated rhythm patterns by the model somewhat diverged from these genres.   

The value of average distance within patterns generated by Creative-GAN is lower than the distances with training data but still relatively higher than the values in Fig.\ref{fig:distance_matrix_cgan}. It means the generated rhythm patterns have some variety, and the model managed to avoid "mode-collapsing." 

\begin{figure}[bt]
  \includegraphics[width=1\columnwidth]{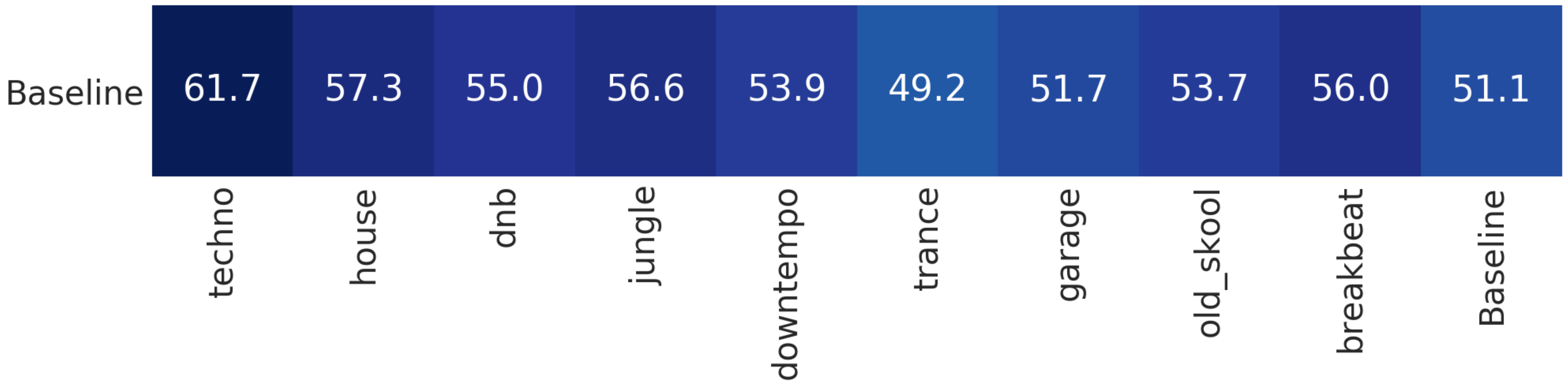}
  \caption{Average distances between randomly generated baseline rhythm patterns and patterns in training data}
  \label{fig:distance_matrix_baseline}
\end{figure}

As a baseline, we have randomly generated drum patterns based on the mean and standard deviation of the number of onsets. Fig.\ref{fig:distance_matrix_baseline} shows higher distance values than Fig.\ref{fig:distance_matrix_cgan}, which means Creative-GAN generated rhythms are closer to training data.

\begin{figure}[bt]
  \includegraphics[width=1\columnwidth]{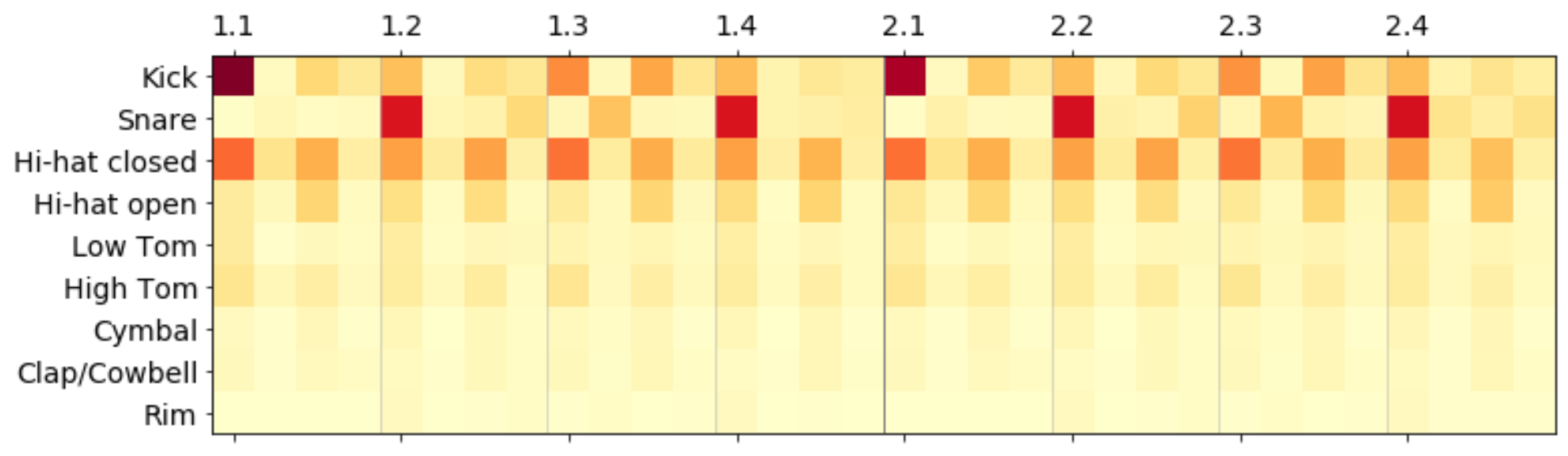}
  \includegraphics[width=1\columnwidth]{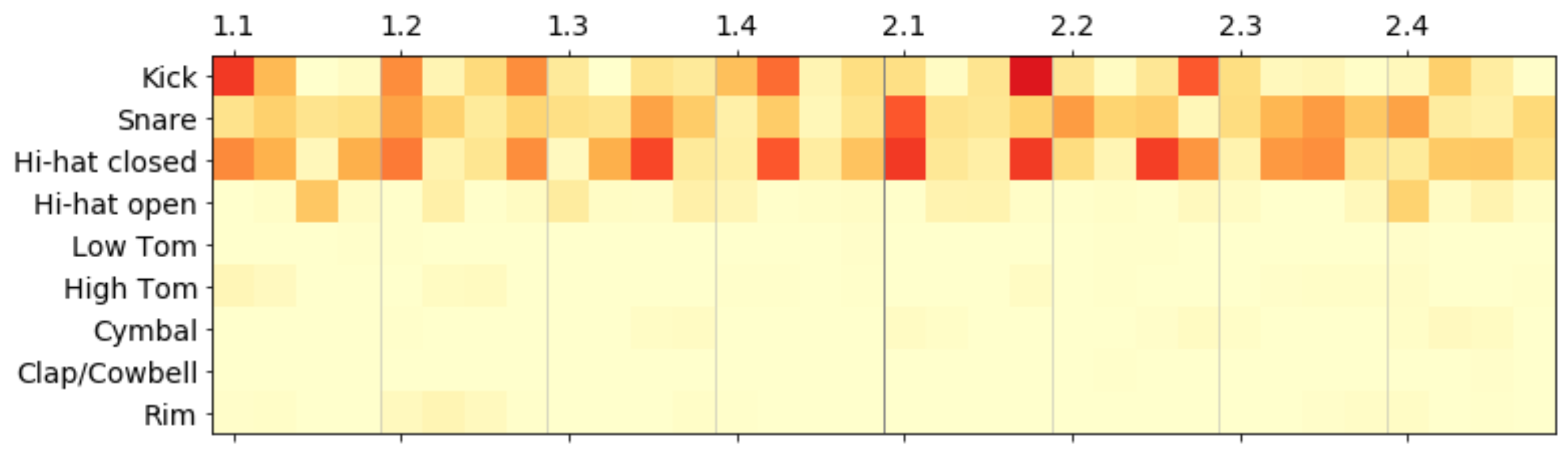}
  \caption{Averaged rhythm onset matrix (top: Training Dataset, bottom: Generated patterns by Creative-GAN model)}
  \label{fig:onset_average}
\end{figure}

Fig.\ref{fig:onset_average} shows the difference between the averaged drum pattern matrices of the training data and generated patterns by the Creative-GAN model.

\section{Discussion}

\subsection{Qualitative analysis}
The result showed that the proposed Creative-GAN model is capable of generating rhythms through adversarial training, and these generated patterns are statistically different from ones in all of the existing genres in training data. 

But a question remains unanswered; {\it are these rhythm patterns really good?}    

The fact that these patterns fooled the discriminator, $D_r$, does not mean they are good as dance music rhythms.  A qualitative analysis of the generated patterns with listening volunteers is yet to be conducted. 

In the case of GAN models for image generation, several qualitative metrics to quantify how realistic generated images are have been proposed, including Inception score(IS)\cite{Salimans2016} and Fréchet inception distance(FID)\cite{Heusel2017}.  Since we aim to generate not only realistic but also unique and creative rhythm patterns, however, such metrics do not apply to our case. It is a challenging task to establish benchmarks to quantify how good these generated data(rhythm patterns) is, even when the data is in an unknown/unique style. 

\subsection{Ableton Live Plugin}

\begin{figure}[bt]
  \includegraphics[width=1\columnwidth]{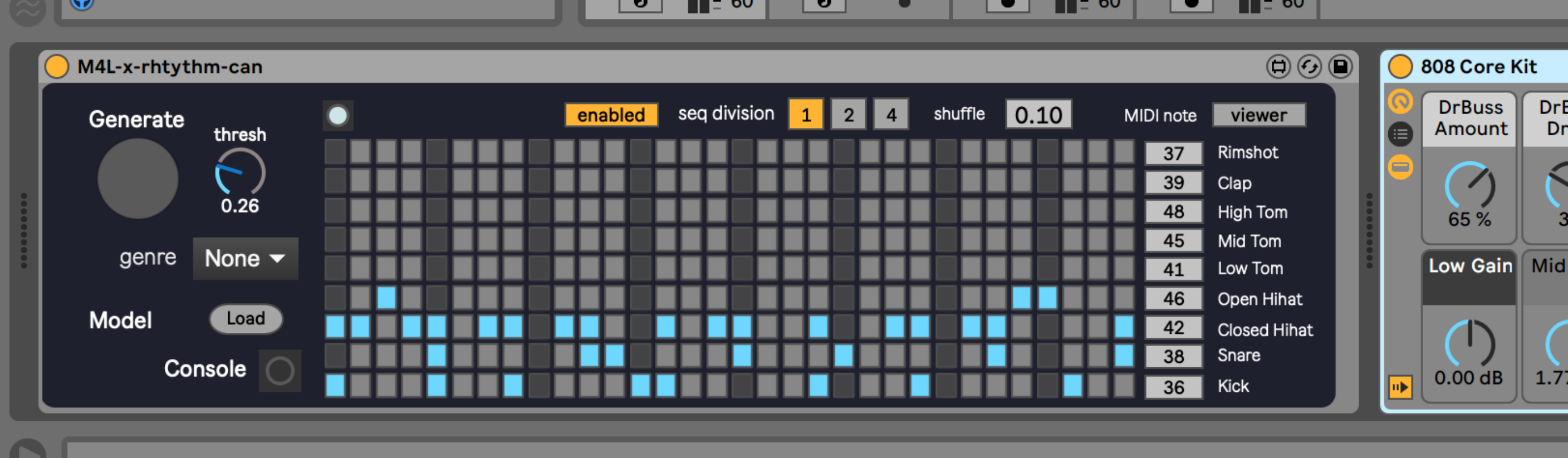}
  \caption{Ableton Live/Max for Live Device}
  \label{m4l_device}
\end{figure}

One of the future directions of this project is to use the technique proposed in this paper in the actual music production of dance music. To help musicians and artists to create new music with the technique described in this paper, the author also provides a plugin software for a popular Digital Audio Workstation(DAW) Software, namely Max for Live environment for Ableton Live (Fig.\ref{m4l_device}). The plugin is implemented in Cycling'74 Max and Node for Max, Node.js environment for Max. The plugin device can load a pre-trained Creative-GAN generator network in TensorFlow.js and generate new rhythm patterns when the user hit the "Generate" button on the plugin. The device can also play generated patterns in sync with the main sequencer of Ableton Live. The design allows users to try various rhythm patterns without being forced to deal with the complicated deep learning environment. The device can also host Genre-conditioned generator networks as well. The plugin is also available on our website\footnote{\url{https://github.com/sfc-computational-creativity-lab/M4L-x-rhythm-can/}}.

\subsection{Micro Timing}

In this paper, the author ignored the micro-timing of the drum onset; every onset was quantized into the timing of 16th notes. 

\cite{Oore2018} and \cite{Gillick2019} showed that the micro-timing---small time offset from the quantized grid-- is essential to realistic music expressions.  Especially in electronic dance music, these offsets strongly contribute to the "grooviness" of the music\cite{sciortino2014would}; For instance, one of the genres in training data, \textit{Garage}, is often characterized by the strong "shuffle" beat. 

We could extend our GAN model to take these micro-timing into account; the generator outputs not only the onset matrix but also micro-timing information, and the discriminator takes timing information as well to determine if they are real or fake and their genres. Our preliminary test showed the genre classification model with timing information worked better.

\section{Conclusion}

Is GAN capable of generating novel rhythm patterns, which do not belong to any of the existing genres, at least genres found in training data?  Can GAN be creative? 

This paper tackled this problem in the realm of electronic dance music, whose various sub-genres tend to be characterized by apparent patterns in the rhythm.

The author has extended the GAN framework by adding the second discriminator. In the training of the discriminator, the second discriminator is trained to classify genres of generated rhythm patterns. On the other hand, the generator is trained adversarially to confuse the discriminator.

The experiment showed that the trained model could generate realistic rhythm patterns in unknown styles, which do not belong to any of the  well established electronic dance music genres in training data.

\section{Acknowledgement}
This research was funded by the Keio Research Institute at SFC Startup Grant and Keio University Gakuji Shinko Shikin grant. The author wishes to thank Stefano Kalonaris and Max Frenzel for inspiration.






\vspace{0.5cm}
\bibliographystyle{iccc}
\bibliography{iccc-can.bib}

\end{document}